# Generalized reciprocal diffractive imaging for stand-alone, reference-free, fast-measurable quantitative phase microscopy

Jeonghun Oh[1,2], Herve Hugonnet[1,2], Weisun Park[1,2], and YongKeun Park[1,2,3,*]

[1]*Department of Physics, Korea Advanced Institute of Science and Technology (KAIST), Daejeon, 34141, Republic of Korea*

[2]*KAIST Institute for Health Science and Technology, Daejeon 34141, Republic of Korea*

[3]*Tomocube, Inc., Daejeon 34051, Republic of Korea*

[*]*yk.park@kaist.ac.kr*

## Abstract

Optical microscopy has been employed to derive salient characteristics of an object in various fields, including cell biology, flow cytometry, biopsy, and neuroscience. In particular, measuring the phase of light scattered from an object aroused great interest by allowing retrieving quantitative parameters such as refractive index, an intrinsic property of a material. Reciprocal diffractive imaging (RDI) has succeeded in recovering the light field scattered from diffusive objects without special restrictions on illumination and sample support from a single-shot intensity in the reference-free regime. However, RDI is limited to imaging samples in the diffusive regime, making application to biological samples difficult. Here, we extend RDI to biological applications by spatially filtering the transmitted fields in the pupil plane. The proposed method is demonstrated by imaging the objects with known structures and various biological samples, showing its capability as a stand-alone optical microscope. We believe that the presented advance could be at the forefront of quantitative phase imaging due to the unique advantages the technique possesses.

**Introduction**

Holography retrieves the whole information of light, that is, amplitude and phase[1], so one can easily manipulate the scattered light field to reconstruct an object of interest. For thin and transparent samples, the retrieved phase information is directly related to the refractive index and the thickness of a sample. Due to this unique characteristic, holographic optical imaging has been exploited for quantitative phase imaging[2], as in histopathology[3-5], microbiology[6-8], hematology[9-13], cell biology[14-17], and preclinical study[18-21]. However, widely adopted holography methods rely on interferometric measurements introducing an additional reference arm, which makes image acquisition vulnerable to ambient noise and requires a bulky optical setup. These hindrances inspired the need for reference-free holographic imaging methods, leading to the development of unhackneyed imaging systems.

The principle of non-interferometric holography fundamentally stands on the close relation between phase and intensity, which is represented by the Siegert relation[22-24]. The transport intensity equation (TIE)[25,26] is one of the non-interferometric techniques and deterministically derives the solution for the phase of light from two intensity measurements: one in-plain and the other defocused. Differential phase microscopy (DPC)[27,28] was proposed to retrieve quantitative wavefronts using two asymmetric illumination patterns. Fourier ptychographic microscopy (FPM)[29,30] reconstructs the light field by iteratively filling the Fourier space from multiple intensity measurements from illumination under various angles. An analytical approach using the Kramers-Kronig (KK) relations[31-34] suggested that under specific sample and illumination conditions, the real part of a complex light field determines the imaginary part, and *vice versa*, and unravels the intertwined relationship between phase and intensity. Quantitative phase imaging using metasurfaces offers significant advantages, including miniaturization, multifunctionality, and compatibility with conventional nanofabrication processes, making it suitable for portable and *in vivo* applications[35,36].

Unfortunately, although the referred methods show promising capability in phase recovery and reveal intriguing physical implications, they have some limitations. The TIE does not provide a universal solution and needs approximations, such as the linear relationship between phase and intensity or strong assumptions for boundary conditions[37]. DPC linearizes the transmittance function of a sample, and KK holographic imaging derives holomorphic properties under the weak object approximation. FPM and KK holographic imaging demand multiple intensity maps to obtain a complete phase image, which hinders the observation of fast dynamics in biological objects. In addition, FPM has a heavy computational cost. It should be noted that non-interferometric methods based on the synthetic aperture approach limit the imaging to thin samples. In addition, the fabrication of metasurfaces remains challenging due to the need for precise nanostructure design and complex manufacturing processes, which can limit their widespread adoption in biomedical imaging systems.

Recently, reciprocal diffractive imaging (RDI)[38] has been proposed for imaging of diffusive objects via non-interferometric detection from a single-shot intensity image. RDI retrieves the complex optical field by reciprocally harnessing the reconstruction principle of coherent diffractive imaging. A Fourier mask that has an asymmetric shape at the pupil plane plays the role of the support in Fienup's hybrid input-output (HIO) algorithm[39], and the constraints in Fourier and image spaces are iteratively applied to the output field in the previous step to reconstruct accurate scattered complex field. This approach provides a unique capability in retrieving diffusive fields, but the application to biological science and more general samples is restricted because the strong components around the DC term are problematic in meeting the condition that the support boundary should be clearly defined.

This study presents a generalization of RDI for imaging samples with the dominant DC components. The Fourier plane of an optical imaging system is modulated by a mask with an irregular shape and a neutral density (ND) filter in order for the support to be well-defined by making near the DC peak weakened. We validate the feasibility of the proposed method for both amplitude and phase objects. Importantly, biological tissues and cells are reconstructed to demonstrate that the proposed method can be utilized in the imaging of various microscopic samples. The proposed RDI method stands out as a self-governing holographic method in a very simple configuration.

**Results**

**Principle**

RDI exploits the HIO algorithm and is a reciprocal version of coherent diffractive imaging[40-42], where the roles of the pupil and image plane are inverted. The measured intensity at the image plane and the known support at the pupil plane are used as constraints. A non-centrosymmetric Fourier mask located at the Fourier plane crops the Fourier field to make it asymmetric; this condition is required for the algorithm to lead to convergence to the correct solution[39,43]. If the Fourier mask is centrosymmetric, the iteration process fails, as shown in Fig. S1. When applying RDI to diffusive objects, the tight edge condition for the support can be easily met because the Fourier spectrum of the scattered field evenly fills the pupil plane. However, in the case of samples not in the diffusive regime, the vicinity of the DC term of a Fourier field generally has a far stronger intensity than the support boundary, which makes the convergence of the HIO algorithm unstable; in this case, the support boundary is not well defined. This issue can be ascertained intuitively or empirically. If the intensity of the part far from the boundary is very strong, the intensity near the boundary will be seen as almost zero, making it difficult to satisfy the 'compact support' condition[44-46]. Figure S2 exhibits how the intensity of the region far from the support boundary affects the reconstruction fidelity. The reconstruction fidelity decreases as the vicinity of the DC term becomes more dominant.

In particular, samples of major interest in biological and medical study, including cells and thin pathological tissue, have a dominant DC term. This means that the previous configuration of RDI cannot be utilized for such samples[38]. To resolve this issue, we modulated the Fourier spectrum by setting an ND filter and an obstruction with a Fourier mask at the Fourier plane to attenuate the intensity of the DC term of the Fourier field. The proposed method was simply implemented in a conventional microscope setup without a reference arm, as shown in Fig. 1a. The only difference with a conventional microscope was that RDI requires locating a Fourier mask at the pupil plane. An ND filter is attached to the Fourier mask to attenuate the low-frequency components (Fig. 1b). The degree of the amplitude attenuation is 1% in this study. This attenuation degree was determined by simulation, considering the effect of noise (see Fig. S3). This Fourier field modulation via the Fourier mask and an ND filter is described with a formula and the specific values of the numerical aperture in Fig. S4. The detailed experimental setup and the design of the mask are presented in the Methods section and Fig. S5.

In addition to diminishing the low spatial frequency components of the Fourier spectrum using an ND filter, criteria for the design of the Fourier mask should be considered: maximizing the reconstruction fidelity while minimizing signal loss. The stronger the strength of the boundary part of the support compared with the interior, the higher the reconstruction fidelity. However, due to the loss of light and vulnerability to noise, the intensity of the low frequencies cannot be lowered indefinitely. To address this issue, we found that cropping near the boundary increases the reconstruction fidelity. Also, it was confirmed through simulation that as the outermost ring part is thinner, the reconstruction fidelity increases. The simulation results depending on the cropping area and the outermost ring part are shown in Figs. S6 and S7. It is presumed that this is because the boundary becomes more dominant compared to the ambient areas. Finally, non-centrosymmetry was ensured by cropping a portion of the support edge. The shape of the used mask was determined as in Fig. 1b. These design criteria are elucidated in Fig. S8.

Figure 2 describes the reconstruction process of the proposed RDI method based on the HIO algorithm. The algorithm reconstructs the sample field using the measured intensity and the information in the Fourier space. This algorithm is a reciprocal version of the HIO algorithm. The sample field to be reconstructed is $s(x, y)$. $S(u, v)$ denotes its two-dimensional (2D) Fourier transform. The sample field is Fourier-transformed and restricted to the shape of the Fourier filter. The intensity of the low-frequency components is greatly reduced according to the ND filter specifications, and only the boundary portion allows light to pass through as it is. The obstruction between the attenuated components and the support boundary helps the algorithm to converge. The modulated Fourier field is denoted by $S'(u,v)$. The intensity of the inverse transform of $S'(u,v)$ gives a constraint in image space:

$$|s'(x,y)| \equiv |\mathcal{F}^{-1}[S'(u,v)]|, \qquad (1)$$

where $\mathcal{F}^{-1}$ denotes the 2D inverse Fourier transform. Using the constraints of the Fourier support and $|s'(x,y)|$ in both Fourier and image spaces, $S'(u,v)$ is retrieved iteratively. The 2D inverse Fourier transform of $S'(u,v)$ is the bandlimited sample field. A detailed description of the HIO algorithm can be found elsewhere[38,39]. We also employed a noise-robust version of the HIO algorithm to maintain robustness in the presence of noise[47]. In simulation, the reconstructed field shows a correlation of 1.0000 with the ground truth, as shown in Fig. 2. Note that the ground truth means the numerical aperture-limited sample field. The generality of the proposed RDI method can be seen by the reconstruction results of various objects in Fig. S9. The reconstruction fidelity depending on the iteration number and the collapsed time depending on the number of reconstructed pixels are described in Fig. S10.

**Experimental validation**

We first verified the proposed RDI method by imaging known samples: a United States Air Force (USAF) resolution target and spherical microbeads as an amplitude object and a phase object, respectively. The measure intensity images showed an unusual distribution because the high-frequency components near the support boundary are transmitted with a much stronger intensity than the rest of the field. The reconstructed amplitude image of the USAF target showed the expected sample structure (Fig. 3a). The smallest pattern of the USAF target (Group 9, Element 3) has 0.78 μm spacing and was well resolved by the imaging system whose theoretical resolution is 0.46 μm. The reconstructed phase image of a microbead shows its expected spherical shapes (Fig. 3b). The phase delay at the bead center was 3.41 rad, and this value is comparable to the results from the off-axis holography setup in Ref. [33].

A more complex bead cluster was also imaged successfully (Fig. 3c).

**Biological application**

Moving to application, we imaged animal tissues (BCN482, US Biolab Corp.) to demonstrate the applicability of the proposed method. Three types of tissue were successfully reconstructed, as shown in Fig. 4. Because these tissues were stained, they also showed contrast in the amplitude images. Both the reconstructed amplitude and phase maps exhibited high contrast and resolution, revealing the appearance of discernable microscopic structures. Because phase unwrapping was broken due to high phase delay in the granule structures, we presented these phase distributions without unwrapping and with a cyclic color map.

Biological cells were also employed in the quantitative microscope. Living biological cells COS7 and Hs68 were imaged, and the reconstructed phase distribution showed the morphology and characteristics of the cells as it is (Figs. 5a and 5b). In addition, a strong point of the RDI method is that this technique only requires a single-shot intensity image, making it an excellent candidate for the imaging of dynamic samples[48,49]. Moving biological cells NIH3T3 were used as subjects to demonstrate the dynamic imaging capability of the proposed method. The phase images of Fig. 5c show the cells change over time with a frame rate of 4.25 fps.

**Discussion**

We proposed RDI for microscopy to retrieve the quantitative phase of various samples. The design of the Fourier mask and the attenuation of the ND filter effectively modulated the transmitted field at the pupil plane of the microscope, enabling the RDI algorithm to work for samples with a strong DC term. The fidelity of the proposed method was demonstrated via objects with known structures. Imaging of biological tissues and cells revealed the promising capability of the proposed method for biological studies.

The RDI method reconstructs complex sample fields from a single-shot intensity measurement in a non-interferometric manner without requiring special optics or a spatial light modulator. In addition, this method does not require oversampling[50] beyond the Nyquist sampling rate, fully utilizing the space-bandwidth product of the imaging system. Although each imaging method has its own advantages, it is interesting to compare the proposed method with the recently developed non-interferometric KK holographic imaging as it has similar advantages. The RDI method only requires a single-shot intensity image and does not exploit an illumination unit. More importantly, the presented method has a significant advantage over KK holography: the reconstruction fidelity is kept as the phase of a sample field increases, as shown in Fig. S11.

A limitation of the proposed method requires further work. Because not only the DC term but also other components of all frequencies we need are attenuated by the ND filter, when the signal-to-noise ratio of the optical system is low, the reconstruction of the frequency components with small signals can be severely affected, as shown in Fig. S3. We anticipate that the use of an apodizing ND filter or a custom-designed ND filter could resolve this issue. Also, the cooperation with developments in methods modulating the Fourier spectra[51] could improve the RDI method for optical microscopy. It is expected that the proposed idea will inspire phase retrieval methods that modulate the Fourier spectrum and take a strategy of artificial intelligence for reconstruction[52-54]. Moreover, variations of gradient descent-based methods would reduce the reconstruction time and iteration number of the RDI method and raise the reconstruction fidelity for noisy situations[55-57].

We believe that the proposed method can easily be applied to optical systems using spatially coherent broadband light sources, such as a superluminescent diode, owing to its non-interferometric characteristic. The use of a temporally incoherent source could suppress coherent noise, expanding applicability to the detection of nano-particles. Furthermore, the RDI method for optical microscopy can be used to reconstruct the 3D refractive index distribution of a sample along with the addition of a module that sweeps the illumination beam using, for example, an LED array[29,58] or a digital micromirror device[59,60]. It is expected that RDI for microscopy will be used not only in biological study[61,62] but also in all areas where holography is utilized, including metrology and inspection[63].

**Methods**

**Experimental setup**

The detailed experimental setup is depicted in Fig. S3. A collimated plane wave illuminates a sample after passing through a tube lens (L1) and a condenser lens (OL1, Olympus Inc., LMPLFLN50X). The light scattered from the sample is collected by an objective lens (OL2, Olympus Inc., UPLSAPO60XW). Then, the light in order passes through relay lenses (L2, L3) and is modulated in the Fourier plane by the custom-made aluminum mask (Fourier mask) and an ND filter (Edmund Optics Inc., 54-459). The modulated light is transferred to a camera (XIMEA, MQ042MG-CM) by a lens (L4). A linear polarizer (Thorlabs, Inc., LPVISE100-A) is located right before the camera. The camera acquires the light intensity. Using the measured intensity distribution and the information for the mask shape, the RDI algorithm retrieves the complex amplitude of the light scattered from the sample.

**Preparation of biological cells**

COS7(ATCC, CRL-1651), Hs68 (ATCC, CRL-1635), and NIH3T3(ATCC, CRL-1658) cells were maintained in Dulbecco's modified Eagle's Medium (DMEM; ATCC, 30-2002) supplemented with 10% fetal bovine serum (Thermo FisherScientific Inc.) and 1% (v/v) penicillin/streptomycin (Thermo Fisher Scientific Inc.) at 37°C in a 5% $CO_2$ incubator. All samples were loaded into imaging dishes (TomoDish, Tomocube Inc.) at a density of $1 \times 10^5$ cells/mL.

## Data and code availability

The MATLAB codes for the implementation of the RDI method and sample data are available at our GitHub repository: github.com/BMOLKAIST/RDI-for-Microscopy-2024.

## Acknowledgements

This work was supported by National Research Foundation of Korea grant funded by the Korea government (MSIT) (RS-2024-00442348, 2022M3H4A1A02074314, RS-2024-00351903), Korea Institute for Advancement of Technology (KIAT) through the International Cooperative R&D program (P0028463), and the Korean Fund for Regenerative Medicine (KFRM) grant funded by the Korea government (the Ministry of Science and ICT and the Ministry of Health & Welfare) (21A0101L1-12).

## Author Contribution

J.O. and Y.P. conceived the project. J.O. designed the optical system, conducted the experiment, and analyzed the data. H.H. provided a discussion of the experimental results. W.P. prepared biological samples. All authors wrote the manuscript. Y.P. provided supervision.

## Competing Interests

The authors declare no competing interests.

**Figures**

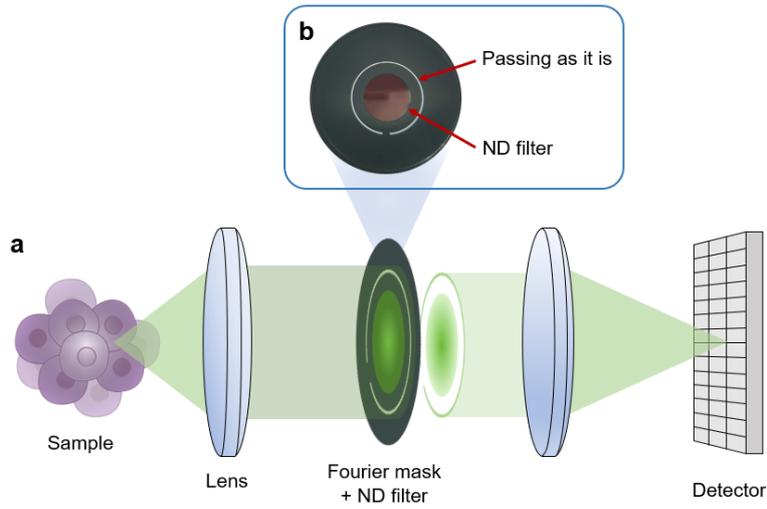

**Fig. 1 | Schematics of reciprocal diffractive imaging (RDI) for general samples.**

(**a**) The scattered field from a sample is transferred to a detector by relay lenses. The combination of the Fourier mask and an ND filter modulates the light in the Fourier plane. (**b**) Photograph of the Fourier mask and the ND filter. The low spatial frequency components in the Fourier spectrum are attenuated by the ND filter, and the light passes through the outermost ring shape.

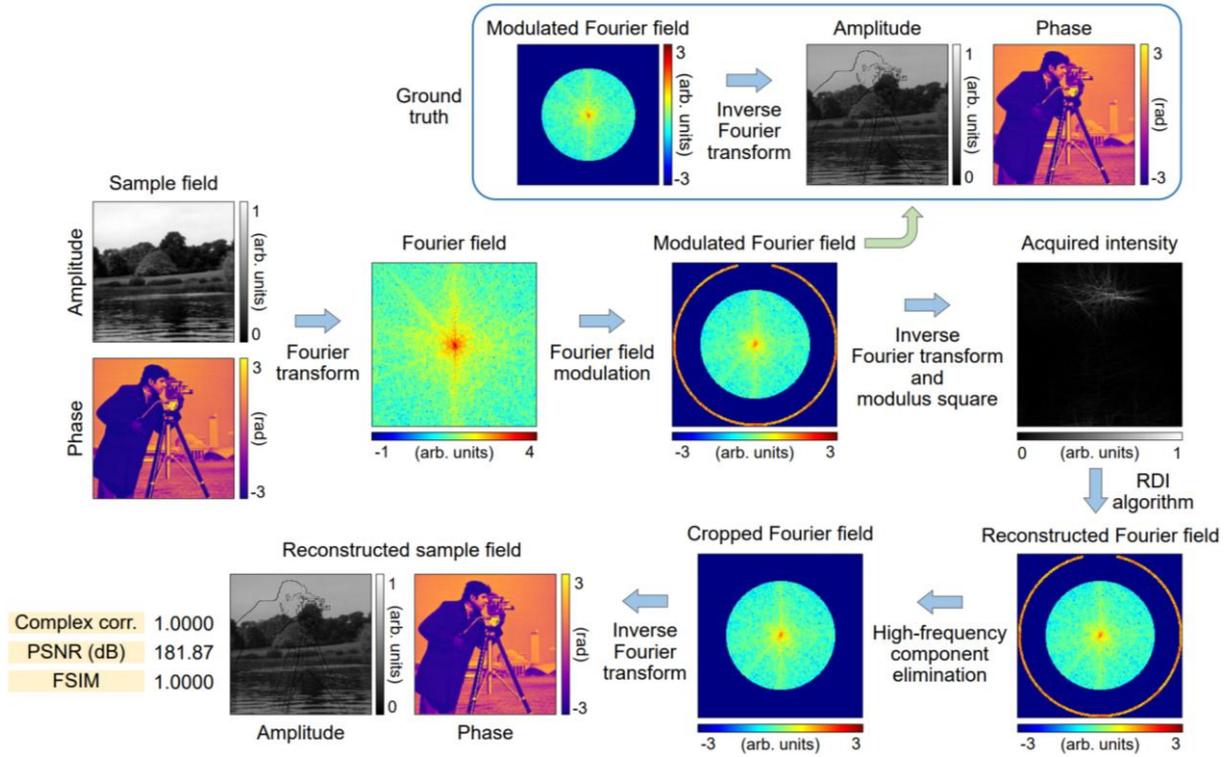

**Fig. 2 | Principle of reciprocal diffractive imaging (RDI) for general samples.**

The 2D Fourier transform of a sample field is modulated in the Fourier plane. The intensity of the 2D inverse Fourier transform of the modulated Fourier field is acquired in a detector. Using the RDI algorithm, the Fourier field is reconstructed. After cropping the outermost ring-shaped part of the Fourier field, its inverse Fourier transform gives the bandlimited sample field.

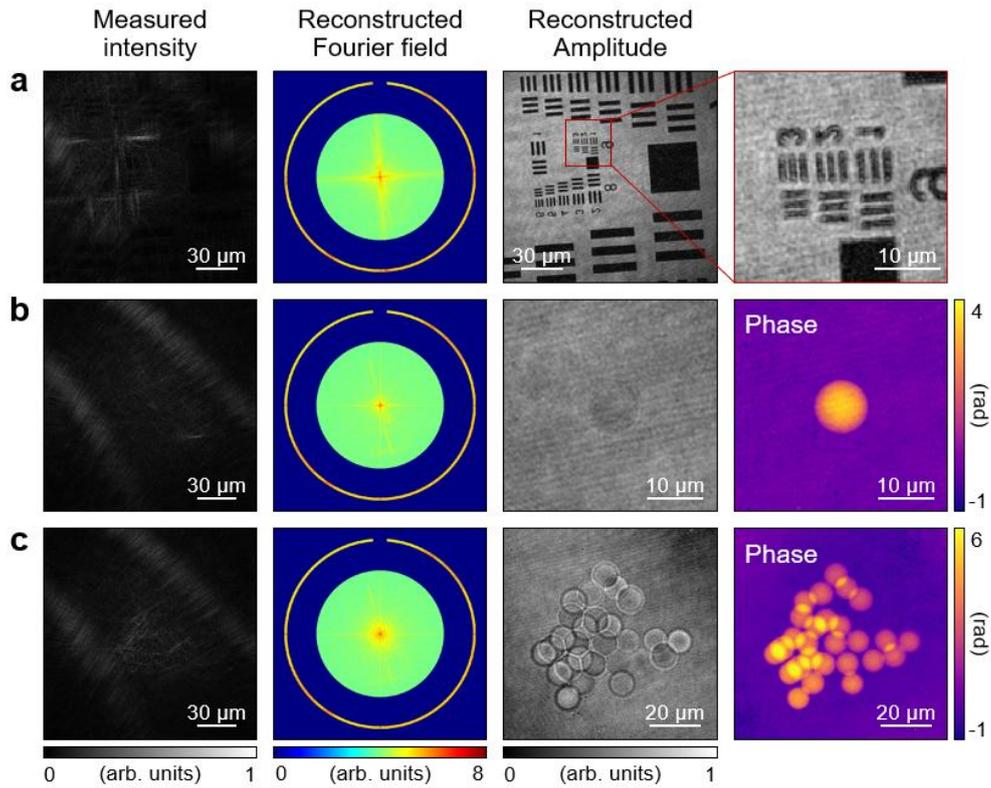

**Fig. 3 | Experimental validation using known samples.**

(**a**–**c**) The sample fields were retrieved by the RDI algorithm: a United States Air Force resolution target (**a**), spherical polystyrene microbead (**b**), and microbead cluster (**c**). The measured intensities, retrieved Fourier fields, retrieved amplitudes, and retrieved phases are shown. The phase unwrapping algorithm was applied. The Fourier spectrum is represented in a log scale.

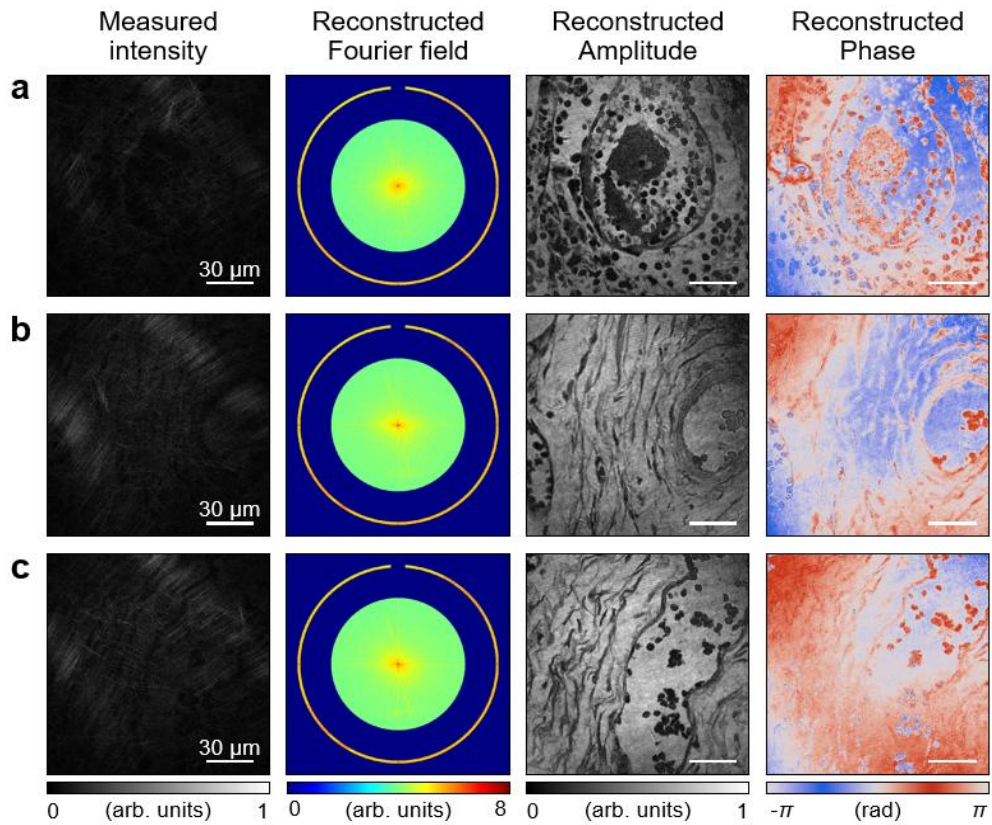

**Fig. 4 | Results for biological tissue imaging.**

(**a**–**c**) The sample fields scattered from biological tissues were retrieved by the RDI algorithm: lymph node (**a**), adjacent normal ovary (**b**), and adjacent normal prostate (**c**) tissues. The measured intensities, retrieved Fourier fields, retrieved amplitudes, and retrieved phases are shown. The Fourier spectrum is represented in a log scale.

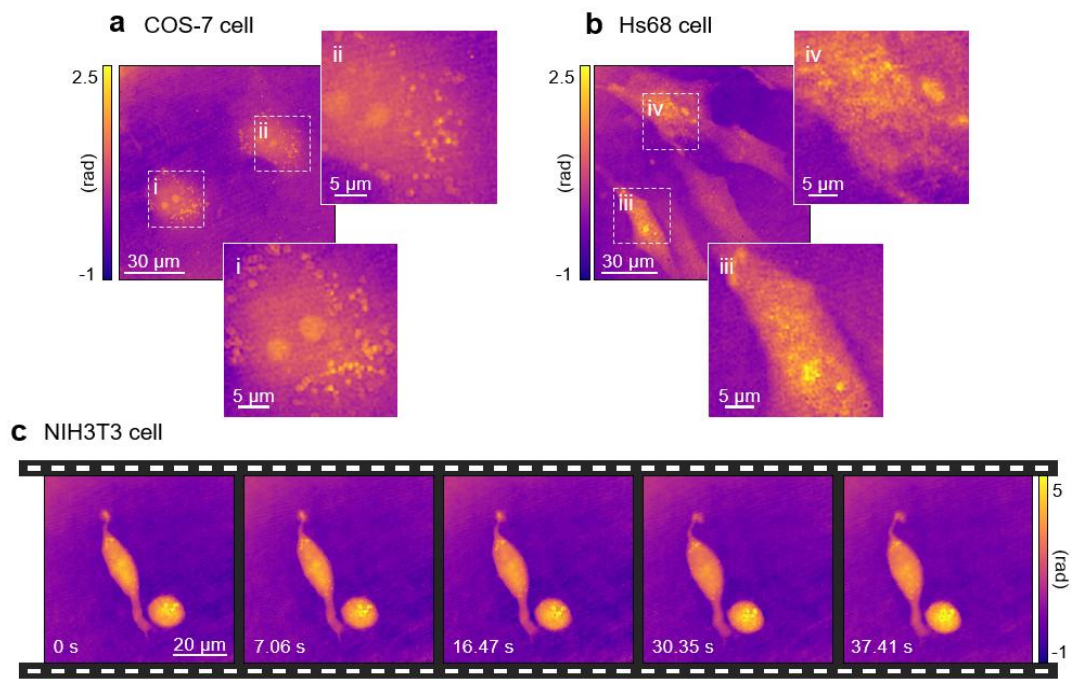

**Fig. 5 | Static and dynamic quantitative phase imaging of biological cells.**

(**a**, **b**) The reconstructed phase distributions of biological cells COS7 (**a**) and Hs68 (**b**). The dotted-line boxes "i–iv" are enlarged to inset for visualization. (**c**) Several frames for the quantitative phase map over time of moving NIH3T3 cells. The phase unwrapping algorithm was applied.